
%
%
%
%
%
%
%
\input phyzzx
%
%
\def\={\!=\!}
\def\-{\!\!-\!\!}
\def\+{\!\!+\!\!}
\def\half{{1\over2}}                
\def\shalf{{\textstyle{1\over2}}}   
\def\d{\delta}
\def\a{\alpha}
\def\<{\left\langle}       
\def\>{\right\rangle}      

\def\e{\epsilon}
\def\la{\lambda}
\def\lh{\hat\lambda}
\def\lt{\tilde\lambda}
\def\dl{{\cal D}\lambda}
\def\r{\rho}
\def\rh{\hat\rho}
\def\dr{{\cal D}\rho}
\def\s{\sigma}
\def\pp{{\rm P}}
\def\N{{\textstyle{1\over N}}}
\def\iN{{\textstyle{i\over N}}}
\def\h{{\cal H}}
\def\T{{\cal T}}
\def\medint{{\textstyle\int}}
\def\medsum{{\textstyle\sum}}
%
%
\def\lhs{\hbox{\it l.h.s.}}
\def\rhs{\hbox{\it r.h.s.}}
%
%
\def\contract#1#2{\vtop{\ialign{##\crcr
   $\hfil\displaystyle{#2}\hfil$\cr\noalign{\kern-6pt} \kern3pt\vrule
   height#1pt\hrulefill \vrule height#1pt\kern3pt\crcr}}}
%
%
\Ref\mehta{ M.L. Mehta: \sl Random matrices\rm, Academic Press
   (New York and London 1967).}
\Ref\revival{ E. Br\'ezin and V.A. Kazakov \journal Phys. Lett.&B236 (90) 144;
\nextline M.R. Douglas and S.H. Shenker \journal Nucl. Phys.&B335 (90) 635;
\nextline D.J. Gross and A.A. Migdal\journal Phys. Rev. Lett.&64 (90) 127,717;
\nextline E. Br\'ezin, M.R. Douglas, V.A. Kazakov and S.H. Shenker
\journal Phys. Lett.&B237 (90) 43; \nextline
   C. Crnkovi\v c, P. Ginsparg and G. Moore \journal Phys. Lett.&B237 (90) 196;
\nextline M.R. Douglas \journal Phys. Lett.&B238 (90) 176.}
\Ref\brezin{ E. Br\'ezin, C. Itzykson, G. Parisi and J.B. Zuber
   \journal Comm. Math. Phys.&59 (78) 35.}
\Ref\bessis{ D. Bessis, C. Itzykson and J.B. Zuber
   \journal Adv. Appl. Math.&1 (80) 109.}
\Ref\sakita{ B. Sakita, \sl Quantum Theory of Many-Variable Systems and Fields
   \rm, World Scientific (Singapore 1985).}
\Ref\jevickijac{ A. Jevicki \journal Nucl. Phys.&B146 (78) 77.}
\Ref\jevickisak{ A. Jevicki and B. Sakita \journal Nucl. Phys.&B185 (81) 89.}
\Ref\cohn{J.D. Cohn and S.P. DeAlwis, IAS preprint IASSNS-HEP-91/7 (Feb.
1991).}
\Ref\karabali{D. Karabali and S. Sakita\journal Int. J. Mod. Phys&A6 (91)
5079.}
\Ref\menahem{ S. Ben-Menahem, SLAC preprint SLAC-PUB-5377 (Nov. 1990).}
\Ref\sohngen{ H. S\"ohngen \journal Math. Z.&45 (39) 245.}
\Ref\tricomi{ F.G. Tricomi \journal Quart. J. Math. Oxford Ser. (2)&2 (51)
199.}
\Ref\titchmarsh{ E.C. Titchmarsh, \sl Introduction to the theory of Fourier
   integrals\rm\ (Oxford, 2nd ed. 1948) Chap.~V.}
\Ref\olaf{ O. Lechtenfeld, IAS preprint IASSNS-HEP-91/2
   (Jan. 1990), to appear in {\sl Journ. Mod. Phys.} {\bf A}.}
%
%
\hfuzz=20pt
\nopubblock
\titlepage
\line{\hfil IASSNS-HEP-91/86}
{\bf\title{Semiclassical Approach to Finite-N Matrix Models\foot{ {\rm
Research supported in part by the Department of Energy, contract
DE-FG02-90ER40542}}}}
\author{Olaf Lechtenfeld\foot{
e-mail: \caps olaf@iassns.bitnet \rm or \caps
lechtenf@guinness.ias.edu}}
\address{School of Natural Sciences
\break Institute for Advanced Study
\break Princeton, NJ 08540}
\vfil
\abstract
We reformulate the zero-dimensional hermitean one-matrix model as a
(nonlocal) collective field theory, for finite~$N$.  The Jacobian
arising by changing variables from matrix eigenvalues to their density
distribution is treated {\it exactly\/}.  The semiclassical loop
expansion turns out {\it not\/} to coincide with the (topological)
${1\over N}$~expansion, because the classical background has a
non-trivial $N$-dependence.  We derive a simple integral equation for
the classical eigenvalue density, which displays strong
non-perturbative behavior around $N\!=\!\infty$.  This leads to IR
singularities in the large-$N$ expansion, but UV divergencies appear
as well, despite remarkable cancellations among the Feynman diagrams.
We evaluate the free energy at the two-loop level and discuss its
regularization.  A simple example serves to illustrate the problems
and admits explicit comparison with orthogonal polynomial results.
\endpage
\chapternumber=0
\pagenumber=1
{\bf\chapter{Introduction}}
Large-$N$ random matrix models have been
with us since the 1950s~[\mehta], and enjoyed a renaissance 1989 due
to newly-found applications to random surfaces, 2d gravity and string
theory~[\revival].  In the late 1970s matrix models encountered gauge
theories, and its two major calculational frameworks were developed:
semiclassical or collective field theory of matrix
eigenvalues~[\brezin], and the technique of orthogonal
polynomials~[\bessis].  The latter has driven the recent successes, by
admitting the investigation of the double-scaling limit and of the
connection to integrable hierarchies.  Given that the semiclassical
approach is particularly simple in the planar limit, it is somewhat
surprising that, except for the one-dimensional model, it has not been
extended to the loop level.  This paper is intended as an attempt in
that direction.

To be sure, there is a well-known obstacle to using the collective
field in a $\N$~expansion of the free energy, namely the exact
evaluation of the Jacobian generated by the change of variables from
matrix eigenvalues to their density distribution [\sakita-\cohn].
However, recent work of Karabali and Sakita~[\karabali] suggests a way
of doing just that.  In this paper we follow their approach and
convert to the Fourier representation of the functional delta-function
defining the eigenvalue density~$\r(x)$.  Its conjugate field,
$\la(x)$, gets thereby introduced into an effective
action~$S_N[\r,\la]$ which, though non-local and non-polynomial, takes
on a fairly simple form.  An equivalent result has been obtained by
Ben-Menahem~[\menahem], through mediating the Coulomb repulsion of the
eigenvalues by a 2d photon field.

In section~2 we derive the effective action and its equation of
motion, for any finite~$N$.  The full saddle-point equation contains
an explicit $\N$~term, which obstructs any attempt for an exact
solution.  At $N\=\infty$, it reduces to the airfoil equation, which
admits exact solutions vanishing outside a compact support~$\s$
[\sohngen,\tricomi,\brezin].  For $N\!<\!\infty$, however, strong
non-perturbative (in~$\N$) behavior appears off~$\s$.  Section~3
presents the semi-classical loop expansion around the classical
background at finite~$N$, \ie\ one-loop determinant, vertices and
propagators.  The Feynman graphs for the free energy~$F_N$ are drawn
at the two-loop level.  The exact but formal expressions are then
expanded around~$N\=\infty$, in section~4.  At first we hide the
implicit $N$-dependence of the saddle-point density~$\rh(x)$ and
obtain remarkably simple results for the loop corrections to~$F_N$,
due to large cancellations among the Feynman diagrams.  We then
develop $\rh(x)$ in~$\N$ as well by iterating a system of airfoil
equations.  The $\N$ expansion is beset by IR and UV singularities
whose regularization is briefly discussed.  Section~5 works out in
detail a simple example and compares to results obtained by orthogonal
polynomial techniques.  The conclusions form section~6, and some
technical material as well as a useful compendium of formulae is
collected in three appendices.
{\bf\chapter{The Classical Eigenvalue Density}}
We begin by
formulating a collective field theory for the zero-dimensional
hermitean one-matrix model at {\it finite\/}~$N$.  Our starting point
is the partition function
$$
Z_N \ \propto\ \int\!d^{N^2}\!\! M \;e^{-N\tr V(M)}
\eqn\ZMdef $$
for an $N\times N$ random hermitean matrix ensemble, in a
potential~$V$.  Upon diagonalization $M=diag(x_i)$ this reduces to
$$
Z_N \ :=\ \biggl[\prod_{i=1}^N \int\! dx_i\biggr] \; \exp\biggl\{
-N\sum_i V(x_i) + \sum_{i<j}\ln (x_i-x_j)^2 \biggr\} \quad.
\eqn\Zxdef $$
We like to change variables from the matrix eigenvalues~$x_i$ to their
density distribution\foot{ Obviously, this transformation is
$N:\aleph_1$ but then $\r$ vanishes almost everywhere.}
$$
\r(x)\ :=\ \N\sum_{i=1}^N\d(x-x_i)\quad.
\eqn\rhodef $$
More precisely, we insert
$$
\eqalign{
1\ &=\ \int\dr\;\prod_x \d\bigl(\r(x)-\N\medsum_i \d(x\-x_i)\bigr)\cr
&=\ \int\!\!\!\!\int\dr\dl\;\exp\Bigl\{i\int\!dx\,\la(x)
\bigl[\r(x)-\N\medsum_i\d(x\-x_i)\bigr]\Bigr\} \cr}
\eqn\insertion $$
into \Zxdef\ and express the action in terms of the density,
$$
S_N^0[\r]\ =\ N^2\int\!\!dx\,\r(x)V(x) -\shalf
N^2\int\!\!\!\!\int\!\!dxdy \;\r(x)f(x-y)\r(y)+\shalf N\,f(0)\quad.
\eqn\Srho $$
The self-interaction had to be regulated by replacing $\,\ln z^2 \to
f(z)\,$ in eq.~\Zxdef, choosing some suitable, \ie\ symmetric and
bounded, function~$f$.

Following ref.~\karabali\ we are able to perform the integration
over~$x_i$,
$$
\eqalign{
Z_N\ &=\ \int\!\!\!\!\int\dr\dl\;\;e^{-S_N^0[\r]+i\int\!\la\,\r}\;
\biggl[\prod_{i=1}^N\int\!dx_i\biggr]\; e^{-\iN\sum_i\la(x_i)} \crr
&=\ \int\!\!\!\!\int\dr\dl\;\;e^{-S_N^0[\r]+i\int\!\la\,\r}\;
\biggl[\int\!dx\;e^{-\iN\la(x)}\biggr]^N \quad,\cr}
\eqn\Zrholambda $$
and arrive at an effective action
$$
S_N[\r,\la]\ =\ S_N^0[\r]-i\int\!dx\,\la(x)\r(x)-N\ln\int\!dx\,e^{-\iN\la(x)}
\eqn\Srholambda $$
which is not only nonlocal in the two real fields $\r$ and~$\la$ but
also non-polynomial in the latter.  Interestingly, the constant mode
of~$\la$ can be integrated out exactly to yield the constraint
$\d(\int\!\r-1)$ that was apparent already from the
definition~\rhodef.  However, we shall keep those modes in the measure
for the time being.  In principle, another constraint arises from the
positivity of~$\r$.  Perturbation theory about a strictly
positive~$\rh$, however, is insensitive to this restriction, and we
will therefore ignore it in the fluctuations.

Our goal is to carry out a systematic semiclassical analysis of this
peculiar one-dimensional field theory.  To leading order in~$\hbar$ we
must determine the saddle-point configurations $(\rh,\lh)$, where the
action~\Srholambda\ is stationary.  The first variations yield
$$
\eqalign{
0\ &=\ N^2 V(x) - N^2\int\!dy\,f(x-y)\,\rh(y) -i\lh(x) \crr
0\ &=\ \rh(x) -e^{-\iN\lh(x)}\Big/\medint\!dy\,e^{-\iN\lh(y)} \cr}
\eqn\SPrholambda $$
where $\rh$ comes out to be properly normalized, $\int\rh=1$.  The
second equation determines~$\lh$ up to a constant,
$$
\lh(x)\ =\ iN\ln\rh(x)+\la_0 \quad,
\eqn\SPlambda $$
a result which may be inserted into the first equation.\foot{ The
first equation subsequently fixes the constant $\la_0$ for a given
solution $\rh$.} Subtracting the value at $x=0$ to get rid of the
constant and deregulating the Coulomb repulsion we get\foot{
See also ref.~[\menahem].}
$$
\bigl[V(x)-V(0)\bigr] - \int\!dy\,\bigl[ \ln(x\-y)^2-\ln y^2 \bigr] \rh(y) +
\N\bigl[ \ln\rh(x)-\ln\rh(0) \bigr] \ =\ 0 \quad.
\eqn\SPrho $$
Together with normalization and positivity, this equation describes
the classical eigenvalue density for any finite~$N$.  It is noteworthy
that~\SPrho\ is not homogeneous in~$\N$, so its solution cannot be,
either.  The saddle-point equation is more easily studied in its
differentiated version,
$$
\shalf V'(x)\ +\ {\textstyle{1\over2N}}{\rh'\over\rh}(x)\ =
\ -\!\!\!\!\!\!\int \! {dy\over x-y}\;\rh(y)\ \equiv\ \pi\,\h_x[\rh]
\eqn\SPrhoc $$
where $-\!\!\!\!\!\int$ denotes Cauchy's principal value of the
integral.  The \rhs\ is known as the Hilbert transform \h\ (of
$\rh$) which has been thoroughly investigated~[\titchmarsh] (see
appendix~A for details).  At this point let us merely state that {\cal
H} is uniquely invertible on the class~$L^p$, $p>1$, by
$$
\h^{-1}_x[\phi]\ =\ -{1\over\pi} -\!\!\!\!\!\!\int\!{dy\over
x-y}\;\phi(y) \quad.
\eqn\Hinv $$

At $N\=\infty$ our equation~\SPrhoc\ has been widely studied and
solved [\sohngen,\tricomi,\brezin], and it was learned~[\olaf] that a
{\it unique\/} solution extends to~$N<\infty$.  Unfortunately, the
equation is not easily solved for finite~$N$.  Even in the large-$N$
limit some care is required, as shown by the following.  For instance,
should we drop the $\rh'/\rh$-term since it is down by~$\N$?  A little
inspection reveals that such a step is in general not consistent with
the asymptotic large-$|x|$ behavior of the equation, which demands a
${1\over x}$ fall-off for the \lhs\ due to the normalization of~$\rh$.
In fact, the formal iterations
$$
\eqalign{
\rh(x)\ &=\ -{1\over\pi^2} -\!\!\!\!\!\!\int\!{dy\over x-y}\;\Bigl[\shalf V'(y)
+{\textstyle{1\over2N}}{\rh'\over\rh}(y)\Bigr] \cr
\rh(x)\ &\sim\ \exp\Bigl\{-N\bigl[V(x)-\medint \ln(x\-y)^2\,\rh(y)\bigr]\Bigr\}
\ \mathrel{\mathop\approx^{|x|\to\infty}}\ x^{2N}\,e^{-NV(x)}\cr}
\eqn\mock $$
show all terms in~\SPrhoc\ to be of the same order for $|x|\gg1$
(unless $V\sim\ln x^2$).  The situation is different, however, near
the minimum of the potential where most of~$\rh$ is
concentrated.\foot{ For simplicity we restrict ourselves to
single-well potentials. See, however, ref.~\olaf.} When $\rh$ is
${\cal O}(1)$ the $\rh'/\rh$-term may safely be neglected or treated
as a $\N$ perturbation in~\SPrhoc.  To the same approximation one
should also drop the exponential tail of~$\rh$ under the integral.  At
$N\=\infty$ this amounts to setting $\rh$ identically to zero beyond a
certain distance from the minimum of~$V$.  The cross-over points $\
x=a,b\ $ between the exterior region (where $\rh$ is ${\cal
O}(e^{-N})$) and the interior domain $\s\equiv(a,b)$ (where it is
${\cal O}(1)$) are determined by the normalization of~$\rh$; they
occur where $\h[\rh]$ can no longer match the growth of~$V'$.
For $N\to\infty$, $\rh\to\r_0$, it is therefore indeed consistent to
simplify eq.~\SPrhoc\ to~[\brezin]
$$
\eqalign{
\shalf V'(x)\ &=\ -\!\!\!\!\!\!\int_a^b \!{dy\over x-y}\,\r_0(y)
\ \equiv\ \pi\,\T_x[\r_0] \qquad {\rm for}\ x\in(a,b)\cr
\r_0(x)\ &=\ 0 \qquad\qquad\qquad\qquad\qquad\qquad\quad
{\rm for}\ x\notin(a,b) \quad.\cr}
\eqn\velocity $$
This relation is known as the {\it airfoil equation\/} and determines
the vorticity~$\r_0$ related to a given velocity field~$V'$ along the
airfoil~$\s$ ($b\-a$ is the span of the wings)~[\sohngen].  It is
solved by inverting the {\it finite\/} Hilbert transform~\T,
$$
\r_0(x)\ =\ {1\over\sqrt{(b-x)(x-a)}}\biggl[\gamma\ -\ {1\over\pi^2}
-\!\!\!\!\!\!\int_a^b \!{dy\over x-y}\;\shalf V'(y)\;\sqrt{(b-y)(y-a)}\biggr]
\eqn\vorticity $$
where the arbitrary constant
$$
\gamma\ =\ {1\over\pi}\int_a^b \! dy\;\r_0(y)
\eqn\constant $$
arises due to the one-dimensional kernel ${1\over\sqrt{(b-x)(x-a)}}$
of~\T\ and is set to $\gamma={1\over\pi}$ by the normalization
of~$\r_0$.  The inversion formula~\vorticity\ is guaranteed to produce
a solution to~\velocity\ as long as $V'\in L^{{4\over3}+\e}$; for
$V'\in L^{1+\e}$ we also need $\r_0\in L^{1+\e}$
in~\vorticity~[\sohngen, \tricomi].  In the case of polynomial
potentials the resulting density~$\r_0(x)$ is seen to vanish at the
endpoints of~$\s$ generically as $(x-a)^{1/2}$ and $(b-x)^{1/2}$.  For
more on the finite Hilbert transform we refer to appendix~A.

Having solved the saddle-point equation in the large-$N$ limit, one is
tempted to try and expand the complete solution $\rh$ of~\SPrhoc\
around~$\r_0$.  A $\N$-expansion appears logical on~$\s$ but is bound
to fail in the exterior region since $e^{-N}$ has an essential
singularity at $N\=\infty$.  Instead, a series in powers of $e^{-NV}$
seems more appropriate there.  Either type of expansion, however, must
break down at the cross-over points, because a smooth function~$\rh$
is to develop from one that is singular there.  For finite~$N$ the
transition from ${\cal O}(1)$ to ${\cal O}(e^{-N})$ occurs in a finite
intermediate region of size $N^{-2/3}$ by a generalized Airy function
of the scaling variable~$(x\-b)N^{2/3}$.  Still, one might hope that a
$\N$-expansion of the classical action needs $\rh$ only on~$\s$, due
to its non-perturbative nature off~$\s$.  We shall pursue this line of
argument in Sect.~4.

It remains to evaluate the classical action in terms of~$\rh$.
Employing eq.~\SPlambda\ as well as $\,\ln\int e^{-\iN\lh}=-\iN\la_0\,$
we find, in agreement with ref.~[\menahem], that
$$
S_N[\rh,\lh]\ =\ N^2\int\!\rh\,V -\shalf N^2\int\!\!\!\!\int\!\rh\,f\,\rh
+N\int\!\rh\,\ln\rh +\shalf Nf(0)
\eqn\SPaction $$
which may be rewritten as
$$
\eqalign{
S_N[\rh,\lh]\ &=\ {N^2\over2}\int\!\rh\,(V\-f) +{N^2\over2}V(0)
+{N\over2}\int\!\rh\,\ln\rh +{N\over2}\ln\rh(0) +{N\over2}f(0) \crr
&=\ {N^2\over2}\int\!\!\!\!\int\rh(x)\bigl[f(x\-y)-f(x)-f(y)\bigr]\rh(y) +
N\ln\rh(0) + {N\over2}f(0) \cr}
\eqn\SPenergy $$
by making use of the saddle-point equation~\SPrho.  Given the
non-trivial $N$-dependence of~$\rh$ it becomes clear that the
semiclassical expansion will not correspond exactly to an expansion
in~$\N$.  On the other hand, it is encouraging to note that
potentially divergent pieces like $\ \int\ln\rh\ $ have dropped out.
The eventually singular $f(0)$~term will get cancelled in the
following section.
{\bf\chapter{Quantum Fluctuations}}
We now set out to develop the loop
expansion of the theory defined by eqs.~\Srho\ and~\Srholambda\ in the
standard fashion.  We eliminate the constant modes from the measure
and split
$$
\r(x)\ =\ \rh(x)+\r'(x) \qquad\qquad \la(x)\ =\ \lh(x)+\la'(x)
\eqn\split $$
with $\int\!\r'=0=\int\!\la'$.  Since $S_N$ is quadratic in~$\r$ its
fluctuations are gaussian and affect only the propagators.  Hence,
only vertices for~$\la'$ are present.  Let us write
$$
Z_N \ =\ Z_{\rm cl}\; Z_{\rm 1-loop}\; Z_+
\qquad\qquad Z_{\rm cl}=e^{-S_N[\rh,\lh]}
\qquad\qquad Z_{\rm 1-loop}=[\det{}' S_N'']^{-\half}
\eqn\Zsplit $$
where $\,\ln Z_+\,$ equals the connected Feynman diagrams, and $\det'$
reminds us to omit the constant modes.  The fluctuation
operator~$S_N''$ at the saddle point reads
$$
S_N''(x,y)\ =\ \pmatrix{-N^2 f(x\-y) & -i\d(x\-y) \cr -i\d(x\-y) &
\N\bigl[\rh(x)\,\d(x\-y)-\rh(x)\rh(y)\bigr] \cr}
\eqn\fluctop $$
and acts on $\{\r'(y),\la'(y)\}$.  The simple form of the off-diagonal
blocks of~$S_N''$ facilitates the calculation of the
determinant,\foot{ $\det\bigl({A\atop-i{\bf1}}{-i{\bf1}\atop
B}\bigr)=\det\,[{\bf1}+A\cdot B]$.  Since $1$ is a zero mode of $B$ we
may omit the primes.}
$$
\eqalign{
Z_{\rm 1-loop}\ &=\ \det{}'\Bigl[\d(x\-y)-N f(x\-y)\rh(y)+
N\medint\!dz\,f(x\-z)\rh(z)\rh(y)\Bigr]^{-\half} \cr
&=\ \exp\biggl\{\sum_{k=1}^\infty {N^k\over2k}\tr\Bigl[f(x\-y)\rh(y)-
\medint\!dz\,f(x\-z)\rh(z)\rh(y)\Bigr]^k\biggr\}\quad,\cr}
\eqn\fluctdet $$
where we have formally expanded in powers of~$N$ (not $\N$).  In the
obvious graphical representation this gives rise not only to the usual
loop diagrams but also to loops ``cut open" at any line, due to the
second term in the trace.  Each term in the above expansion seems to
be well-behaved except for the very first one, ${N\over2}\tr
f\rh={N\over2}f(0)$, which in fact just cancels the eventually
singular term in the classical action, eq.~\SPaction.  Of course, one
would rather like to expand in powers of~$\N$ but let us defer this to
the next section.

Instead, we complete the construction of the Feynman rules.  To obtain
the $\la'$ propagator we need to invert~$S_N''$,\foot{ The inversion
of $f$ will be discussed in the next section.}
$$
\eqalign{
\contract{12}{\la'(x)\;\la'}(y)\ &=\ N\Bigl[\bigl(\rh(x)\d(x\-y)-\rh(x)\rh(y)
\bigr)-\N f^{-1}(x,y)\Bigr]^{-1}\cr
&=-N^2 f(x,y)-N^3\!\int\!\!\!\!\int f(x,z)
\bigl(\rh(z)\d(z\-w)-\rh(z)\rh(w)\bigr) f(w,y)-\ldots \cr}
\eqn\propagator $$
again formally expanding in~$N$.  The corresponding graphs are the
same as in~\fluctdet\ before taking the trace.

Turning to the vertices, the $\la'$~expansion simplifies at the saddle
point as
$$
\eqalign{
\Delta S\ &\equiv\ -N\ln\int\!dx\,e^{-\iN\la(x)} \ =
\ i\la_0 -N\ln\int\!\!dx\,\rh(x)\,e^{-\iN\la'(x)}\crr
&=\ i\la_0 +N\sum_{r=1}^\infty {(-1)^r\over r}\biggl[\sum_{s=1}^\infty
{1\over s!}\Bigl({-i\over N}\Bigr)^s\,\int\!\!dx\,\rh(x)\la'(x)^s\biggr]^r \cr}
\eqn\lambdaexp $$
with the help of $\,e^{-\iN\lh}\big/\int e^{-\iN\lh}=\rh$.  The
constant mode of~$\la'$ appears only at ${\cal O}(1)$, as it should.
Note that this is a genuine $\N$~expansion because in powers
of~$\iN\la'$.  The ${\cal O}(1)$ and ${\cal O}(\N)$ terms have already
been taken care of in~\SPrholambda\ and~\fluctop.

It is instructive to compute the two-loop correction to the free
energy
$$
F_N\ =\ -\ln Z_N\ =\ S_N[\rh,\lh] + \shalf\tr'\ln S_N'' -\ln Z_+ \quad.
\eqn\freeenergy $$
Writing $\Delta S=N\sum_k N^{-k}S_k$, with $k$ lines emanating
from~$S_k$, we have
$$
Z_+\ =\ \<e^{-S_3/N^2-S_4/N^3-\ldots}\>
\ =\ \exp\Bigl\{-N^{-3}\VEV{S_4} +\half N^{-4}\VEV{S_3 S_3} -\ldots\Bigr\}
\eqn\Ztwoloop $$
where $\VEV{\ldots}$ denotes the free field average performed with the
propagator~\propagator.  With a little combinatorics $S_k$ can be read
off eq.~\lambdaexp, \eg\
$$
\eqalign{
S_3\ &=\ {(-i)^3\over3!}\Bigl[-\medint\rh\la'^3
+3\medint\rh\la'^2\!\medint\rh\la' -2(\medint\rh\la')^3\Bigr] \cr S_4
\ &=\ {(-i)^4\over4!}\Bigl[-\medint\rh\la'^4
+4\medint\rh\la'^3\!\medint\rh\la' +3(\!\medint\rh\la'^2)^2
-12\medint\rh\la'^2(\!\medint\rh\la')^2 + 6(\!\medint\rh\la')^4\Bigr]\cr}
\eqn\Sk $$
which are depicted in figure~1.
Obviously, the parts of~$S_k$ are labelled by the partitions of the number~$k$
into integers.
The $\la'$ contractions produce 21 diagrams which are displayed in
figure~2.  Due to the non-locality of the vertices, the terms ``loop"
and ``vertex" take on {\it two\/} distinct meanings, namely one
counting powers of~$\hbar$ and denoting~$S_k$, and the other
describing individual diagrams.  In order to avoid notational
confusion, we distinguish between the two and refer to figure~2 as the
two-loop {\it contributions\/} built from {\it vertices\/} $S_k$,
represented by 6 two-loop, 9 one-loop, and 6 tree-level {\it
diagrams\/} consisting of 2 or 3 lines and up to 6 {\it nodes\/}.
Note also that all contributions to $\ln Z_+$ are connected whereas
the constituting graphs may not be.
It appears that we have derived a sensible set of Feynman rules
\fluctdet-\lambdaexp\ for our non-local theory.

The functional $\Delta S$ in~\lambdaexp\ has the property that
an $x$-independent shift of $\la'$ leaves the vertices~$S_k$ form-invariant.
We may exploit this fact to simplify the structure of~$S_k$,
by changing the field variable $\la'\to\lt:=\la'-\int\!\rh\la'$.\foot{
The Jacobian of this transformation is compensated by the change in
the one-loop determinant.}
Since $\int\!\rh\lt=0$, a large number of terms disappear when $S_k$ are
expressed in terms of~$\lt$.
We are left with the parts corresponding to partitions of~$k$ into
integers~$\ge2$ only.
The 21~diagrams of figure~2 compactify to the five graphs without
univalent nodes.
In particular, tree graphs can no longer be present.
The price we pay for this simplification is a more involved propagator
for~$\lt$, which actually consists of four terms when rewritten
in terms of the $\la'$ propagator.

We gain, however, an argument showing the cancellation of all
tadpole infinities.
The only potentially divergent diagrams arise from self-contractions,
$$
\contract{12}{\la'(x)\,\la'}(x) \ =\ -N^2\,f(0)\ +\ {\rm finite\ terms} \quad.
\eqn\self $$
Any $\lt$-propagator ending in a univalent node with a weight of
$constant\times\rh$ is seen to vanish,
which eliminates all $f(0)$ terms from tadpoles.
The singularities arising from self-contracted multiply valent nodes appear to
cancel systematically between different graphs (see figure~2 for example).
This argument holds true for any {\it constant\/}
part of $\contract{12}{\la'(x)\,\la'}(x)$, whether we expand
in~$N$ or not.

Expanding the full propagators in the loop graphs \`a la~\propagator\
leads to an infinite series in powers of~$N$, which ruins the $\N$
expansion from~\lambdaexp.  Indeed, absorbing the leading
$N$-dependence of $\contract{12}{\la'(x)\,\la'}(y)$ into the vertices
in~\Ztwoloop\ replaces $S_k/N^{k-1}\to N S_k$, for any~$k$, so that
all terms in $\Delta S$ contribute at lowest order in~$N$.  It
therefore appears that we need to sum the loop expansion in order to
consistently develop $F_N$ in powers of~$N$.  On the other hand, $F_N$
has a logarithmic singularity at $N\=0$ (at least for a quadratic
matrix potential) which makes an $N$ expansion ill-defined anyway.

Nevertheless, the semiclassical expansion seems to be well-behaved for
finite~$N$.  It is merely dangerous to develop the individual loop
contributions in~$N$ (or $\N$ as we shall see soon).  Unfortunately,
for practical purposes we need to compute the determinant~\fluctdet\
and the propagator~\propagator\ somewhere, and this seems feasible
only at the extrema $N\=0$ or $N\=\infty$.  Another option is to put
the system in a box (in eigenvalue space) of length~$L$ and estimate
determinant and propagator in Fourier space.  We shall come back to
this approach later on.
{\bf\chapter{$\bf{1\over N}$ Expansion}}
This section is devoted to an
attempt to set up a systematic $\N$ expansion of the free
energy~\freeenergy, using the previous results \SPrhoc-\SPenergy\ and
\fluctdet-\lambdaexp.  We will proceed in two stages.  First, we
arrange according to explicit powers of~$\N$ and hide the implicit
$N$-dependence of~$\rh$.  Second, we expand $\rh$ as well and insert
the result into the coefficients obtained earlier.

The classical part of the free energy given by eqs.~\SPaction\
or~\SPenergy, it remains to develop the fluctuation
determinant~\fluctdet\ and the full $\la'$~propagator~\propagator\
around $N\=\infty$.  To this end one needs to invert $\
NS_N''|_{\la\la}=\rh(x)\d(x\-y)-\rh(x)\rh(y)\ $ away from the constant
mode.  As expected, this operator has a constant zero mode since
$\int\!\rh=1$.  Hence, its inverse orthogonal to that mode is
simply\foot{ It is convenient to make a change of basis,
$\la'=\xi/\sqrt{\rh}$, so that the kernel of
$\d(x\-y)-\sqrt{\rh(x)\rh(y)}$ is spanned by~$\sqrt{\rh}$.  Because
$\sqrt{\rh(x)\rh(y)}$ projects onto the kernel, our operator projects
onto the complement and is inverted by the identity there.}
$$
\bigl[\rh(x)\d(x\-y)-\rh(x)\rh(y)\bigr]_\bot^{-1}\ =
\ {\textstyle{1\over\rh(x)}}\,\d(x-y) \quad.
\eqn\inverse $$
On the level of the determinant this can also be seen in Fourier
space, from the identity
$$
\det{}'[\rh_{k+l}-\rh_k\,\rh_l]\ =\ \det[\rh_{k+l}]
\qquad {\rm for}\quad \rh_0=1\quad.
\eqn\modedet $$
The propagator~\propagator\ is then easily written out as
$$
\contract{12}{\la'(x)\;\la'}(y)\ =
\ {\textstyle N {1\over\rh(x)}\,\d(x\-y)
\ +\ {1\over\rh(x)}f^{-1}(x,y){1\over\rh(y)}
\ +\ \N\int{1\over\rh}\,f^{-1}\,{1\over\rh}\,f^{-1}\,{1\over\rh}
\ +\ {\cal O}({1\over N^2}) } \,.
\eqn\propexpansion $$
The one-loop determinant~\fluctdet\ works out analogously,
$$
\eqalign{
Z_{\rm 1-loop}\ &=\ \det\Bigl[-N f(x\-y)\rh(y)\Bigr]^{-\half}\times
\det\Bigl[\d(x\-y)-\N{\textstyle{1\over\rh(x)}}f^{-1}(x,y)\Bigr]^{-\half}\crr
&\propto\ \exp\biggl\{-\shalf \eta\ln N-\shalf\!\int\!\ln\rh\biggr\}
\ \times\ \exp\biggl\{\sum_{k=1}^\infty {N^{-k}\over2k}
\tr\Bigl[\rh(x)^{-1}f^{-1}(x,y)\Bigr]^k\biggr\} \cr}
\eqn\detexpansion $$
where $\eta$ is a regulating constant, and we dropped a singular
constant determinant.  Together with eqs.~\SPenergy\ and~\lambdaexp\
this yields a systematic but formal expansion of the free
energy~\freeenergy.  Each $k$-loop contribution starts at~${\cal
O}(N^{-k+1})$ and is made of $k$- and lesser-loop graphs, many of
which are disconnected.  The ensuing Feynman rules are given in
figure~3, but the combinatorical coefficients and the occurrence of
the nodes have to be read off the expansions~\lambdaexp,
\propexpansion\ and~\detexpansion.  Clearly, the expansion is not just
in~$\N$ but also in~${1\over N\rh}$, which will be responsible for IR
divergencies later on.

We have yet to give a less formal meaning to the
``propagator"~$f^{-1}$.
Since the convolution with
${1\over2\pi}\partial_x f(x,y)\to{1\over\pi}\pp{1\over x-y}$ is just
the Hilbert transform~\SPrhoc, the deregulated $f(x\-y)\to\ln(x\-y)^2$
is easily inverted from~\Hinv,\foot{ The symbol $\pp$ denotes the
principal value, $\int\pp\ldots=-\!\!\!\!\!\!\int\ldots$, or $\
\pp{1\over x}={1\over x}-i\pi\d(x)$.}
$$
f^{-1}(x,y)\ \to\
-{1\over2\pi^2}\,\pp{1\over x-y}\,{d\over dy} \ =\
-{1\over2\pi^2}\,{d\over dx}\,\pp{1\over x-y}
\eqn\finv $$
on the space $L^{1+\e}$.

Employing the results of appendix~A we are now able to compute the
one-loop traces of eq.~\detexpansion.
The derivatives get distributed
over the graph, and, after evaluating
$$
\eqalign{
\int\!\!\!\!\int\!{\pp\over y-x}\phi_1(x){\pp\over x-y}\phi_2(y)&=
\int\!\h'[\phi_1]\,\phi_2 -\d(0)\int\!\phi_1\,\phi_2 \cr
\int\!\!\!\!\int\!\!\!\!\int\!
{\pp\over z-x}\phi_1(x){\pp\over x-y}\phi_2(y){\pp\over y-z}\phi_3(z)&=
\int\bigl(\h'[\phi_1]\,\phi_2 - \phi_1\,\h'[\phi_2]\bigr)\h[\phi_3] \cr}
\eqn\ploops $$
and so on, we discover that $\tr\bigl[\rh^{-1}f^{-1}\bigr]^k$ vanish
upon symmetrization over cyclic permutations!\foot{
We have checked this for $k\le3$.}
Hence, the one-loop correction to $F_N$ comes from
the first (singular) factor of~\detexpansion\ only and is given by
$$
\shalf\tr{}'\ln S_N''\ =\ \shalf \eta\ln N\ +\ \shalf\int\!dx\,\ln\rh(x) \quad.
\eqn\oneloop $$
By the same calculation, self-contractions
$\contract{12}{\la'(x)\,\la'}(x)$ do not receive $\N$ corrections
beyond the leading term.

To order $N^{-1}$ we have to review our two-loop calculation of the
previous section.  It suffices to replace all full propagators of
figure~2 by $\ N\rh(x)^{-1}\d(x\-y)$.  In terms of the Feynman rules
of figure~3 this means that we compose our diagrams with dashed lines
only.  Thus any propagator loop yields a singular factor of~$\d(0)$,
leaving us with connected graph components proportional to
$\d(0)^2\rh^{-1}$, $\d(0)$, or~$\rh$.  It is quite remarkable that
all but one of the divergent graphs cancel between $\VEV{S_4}$
and~$\VEV{S_3\,S_3}$!\foot{ The propagator self-contraction
cancellation argument of the previous section does not even apply
since $\contract{8}{\la'(x)\,\la'}(x)=N\d(0)/\rh(x)$ is not constant.}
However, the single genuine two-loop diagram survives (together with
some finite contributions),
$$
-\ln Z_+\ =\ \biggl[{1\over12}\,\d(0)^2
\int{dx\over\rh(x)}\ -\ {1\over12}\biggr]{1\over N}
\ +\ {\cal O}({1\over N^2})\quad.
\eqn\twoloop $$

It is then quite clear that a $k$-loop contribution to~$\ln Z_+$ will
yield (products of) expressions of the form $\d(0)^p \int\rh^{1-p}$,
with $0\le p\le k$.  However, all Feynman graphs appear to cancel each
other except for those with $p\=0$ and $p\=k$.  The latter are
simply given by all genuine (connected) $k$-loop graphs, presumably without
self-contractions.\foot{ We have checked
this also on the 3-loop level, with an outcome of
${1\over12}\d(0)^3\int(\rh N)^{-2}+\hbox{\rm finite}$.}

It should not have escaped the attentive reader that the
expressions~\oneloop\ and~\twoloop\ are divergent.  In fact, we
encounter two different types of singularities.  First, the
regulator~$\eta$ as well as the factors of~$\d(0)$ arose from
evaluating $\tr{\bf1}$ and thus count the (infinite) number of degrees
of freedom.  Second, the integrals in both expressions are singular
since $\rh\sim e^{-NV}$ for $|x|\to\infty$ induces $\int\ln\rh\sim
N\int V$ and $\int\rh^{-1}\sim\int e^{+NV}$.  The latter infinities
require an IR cutoff; most convenient is a restriction of the
eigenvalue range to a finite interval, say
$x\in[-{L\over2},+{L\over2}]$, with open boundary conditions.  Putting
the system into a box does not, however, cure the UV singularities
mentioned first; those ask for a mode cutoff, say $|k|\le M$, for some
large $M$.  A natural choice for the value of~$M$ would seem $M\=\a
N\+\beta$.  We have estimated the resulting leading behavior of the
one-loop determinant~\fluctdet, with some details given in appendix~C.
The upshot is
$$
\shalf\tr{}'\ln S''_N\Big|_{\rm reg}\ =
\ \bigl[(\a\+1)\ln(\a\+1)-\a\ln\a\bigr]N +\ {\cal O}(\ln N)\quad.
\eqn\estimate $$
The ${\cal O}(N)$ term originates from $\half\ln\det[-f]$ which we
dropped in~\detexpansion.  Since we did not probe the $\rh$
dependence, $L$ does not appear in our result.  Further numerical
analysis is required to elucidate the cut-off dependence of the
regulated loop corrections.

The coefficients of the $\N$ series we have obtained are (singular)
functionals of the saddle-point density~$\rh$.  To complete the $\N$
expansion we finally develop $\rh$ about $N\=\infty$ as well, thus
blinding ourselves to its ``non-perturbative" parts.  In $\N$
perturbation we set $\rh(x)\equiv0$ for $x\notin\s$ and expand
$$
\rh(x)\ =\ \sum_{i=0}^\infty N^{-i}\,\r_i(x) \qquad {\rm for\ } x\in\s \quad.
\eqn\rhosplit $$
Next, we define a sequence of potentials
$$
V_0=V\quad,\qquad
V_1=\ln\r_0\quad,\qquad
\sum_{i=1}^\infty N^{-i} V_{i+1}=\sum_{r=1}^\infty {(-)^{r-1}\over r}
\biggl(\sum_{s=1}^\infty N^{-s}{\r_s\over\r_0}\biggr)^r
\eqn\pot $$
by expanding $\ln\rh$.  The saddle-point equation~\SPrhoc\ then turns
into the iteration
$$
\shalf V'_i(x)\ =\ -\!\!\!\!\!\!\int_a^b\!{dy\over x-y}\;\r_i(y)
\qquad {\rm for\ } x\in(a,b)
\eqn\iteration $$
since $V_i$ is a function of $\{\r_0,\r_1,\ldots,\r_{i-1}\}$.  The
constant part of $\r_i$ is fixed by the normalization conditions
$\int\!\r_i=\d_{i,0}$.  Clearly, the iteration consists of repeated
use of eqn.~\vorticity\ for $\T^{-1}$, with $\gamma\=0$ after
the first step.

Having computed the coefficients in~\rhosplit\ to a desired order
in~$\N$, we are able to unravel the complete large-$N$ expansion of
the free energy.  First, the above equations can be employed to
simplify the $\N$~series of the classical action~\SPenergy:
$$
\eqalign{
S_N[\rh,\lh]&=\ {N^2\over2}\!\int\!\!\r_0\bigl[V(x)\-V(0)\-f(x)\bigr]
+{N\over2}f(0)+N\!\int\!\!\r_0\ln\r_0+\half\int\!\!\r_1\ln\r_0+\ldots\crr
&=\ \shalf N^2\!\int\!\!\!\!\int\r_0(x)\bigl[f(x\-y)\-f(x)\-f(y)\bigr]\r_0(y)
+\shalf N f(0)+N\ln\r_0(0)\cr
&\quad +N\!\int\!\!\!\!\int\!\r_0(x)\bigl[f(x\-y)\-f(y)\bigr]\r_1(y)
+\shalf\!\int\!\!\!\!\int\!\r_1(x)f(x\-y)\r_1(y)+{\cal O}(\N)\;.\cr}
\eqn\SPexpansion $$
Second, we insert the expansion~\rhosplit\ into the
expressions~\oneloop\ and~\twoloop\ for the loop corrections.
However, by setting $\rh\!\sim\!e^{-NV}$ to zero off~$\s$ we are
dropping infinite contributions $\sim\!N\!\int\!V$ and
$\sim\!\int\!e^{+NV}$ to the integrals of~\oneloop\ and~\twoloop,
respectively, of which the first is perfectly perturbative.

At this stage it is quite obvious that a brute-force expansion around
$N\=\infty$ makes no sense.  In view of the interference with strong
$e^{-N}$ behavior one must keep at least the $N$-dependence of~$\rh$
intact.  Only then may one hope to establish a large-$N$ series for
the free energy, provided the regulation problems can be overcome.
Under these circumstances it is hard to see how a serious comparison
with orthogonal polynomial results could be achieved beyond the
leading order.
{\bf\chapter{An Example}}
We have tried to develop the semiclassical
expansion for the zero-dimensional hermitean matrix model with an
arbitrary potential~$V$.  In order to explicitly demonstrate the
difficulties of this scheme we now become more specific and choose a
simple potential,
$$
V(x) \ =\ \shalf x^2 + g\,x^4 \quad,\qquad g>0 \quad,
\eqn\quartic $$
for which we can evaluate the free energy~$F_N$ by the method of
orthogonal polynomials.  From the literature~[\bessis] we extract the
value for $F_N(g)\-F_N(0)$, and add to it the result for the quadratic
potential ($g\=0$), as calculated in appendix~B.  The first orders in
$\N$ yield
$$
\eqalign{
F_N\
&=\ \Bigl[{3\over4}+{1\over24}(a^2-1)(9-a^2)-\half\ln a^2\Bigr]N^2-N\ln N\cr
&\quad + \bigl[1-\ln2\pi\bigr] N - {5\over12}\ln N +
\Bigl[c_0 +{1\over12}\ln(2-a^2)\Bigr] + {\cal O}(N^{-2})\cr}
\eqn\OPresult $$
with
$$
12g a^4 + a^2 -1 =0 \qquad\Longleftrightarrow\qquad a^2 =
\bigl(-1+\sqrt{1+48g}\bigr)\big/24g
\eqn\adef $$
and $c_0\approx-0.7535$, a universal constant.

Let us see how much of this the collective field is able to reproduce.
The iterative solution of eqs.~\pot\ and~\iteration\ proceeds along
standard lines.  To lowest order eq.~\vorticity\ produces~[\brezin]
$$
\r_0(x)\ =\ {\textstyle{1\over2\pi}}\bigl(1+8ga^2+4gx^2\bigr)\sqrt{4a^2-x^2}
\;\Theta(4a^2-x^2)\quad,
\eqn\rhozero $$
a polynomial deformation of Wigner's semicircle law.  The next step in
the iteration~\pot\ and~\iteration\ yields
$$
V'_1(x)\ =\ {2x\over x^2+n^2} -{x\over4a^2-x^2}\qquad {\rm with} \quad
n^2={\textstyle{1+8ga^2\over4g}={2+a^2\over1-a^2}a^2}
\eqn\Vone $$
which is to be plugged into~\vorticity\ to find the
correction~$\r_1=\r_1^{(1)}\+\r_1^{(0)}$.  With the help of
appendix~A\foot{ It it useful to remember that
${1\over\pi}-\!\!\!\!\!\!\int_{-1}^{+1}\!dx\,{\sqrt{1-x^2}\over
x^2\pm\a^2} =-1+\Theta(\a^2\-1)\,\sqrt{1\pm\a^{-2}}$.} the first term
gives
$$
\r_1^{(1)}(x)\ =\ {1\over\pi}{1\over\sqrt{4a^2-x^2}}\biggl[
{\sqrt{(4a^2+n^2)\,n^2}\over x^2+n^2}\ -1 \biggr]\;\Theta(4a^2-x^2)
\eqn\rhoonereg $$
but the second term is not integrable at $x=\pm2a$; stretching the
meaning of~\iteration\ we might set
$$
\r_1^{(0)}(x)\ =\ {1\over4}\bigl[\d(x-2a)+\d(x+2a)\bigr]
-{1\over4\pi}{\Theta(4a^2-x^2)\over\sqrt{4a^2-x^2}} \quad.
\eqn\rhoonesing $$
These results continue to hold true for $-{1\over48}<g<0$; in this
regime $2>a^2>1$ and $-8>n^2>-\infty$.  At the lower bound $|n|\to2a$,
and the well-known additional zero develops at the edge of~$\s$.
Evidently, the iteration~\iteration\ leads to ever worse singularities
at~$x=\pm2a$.  This merely reflects the breakdown of the
$\N$~expansion at the edges and shows that our neglect of the
non-perturbative tail of~$\rh$ was illegitimate.

Nevertheless, we insert the above findings into the
functionals~\SPexpansion,
\oneloop\ and~\twoloop.
The classical action becomes
$$
\eqalign{
S_N[\rh,\lh]\ &=\ N^2\Bigl[{3\over4}+{1\over24}(a^2-1)(9-a^2)-\half\ln
a^2\Bigr]
\ +\ \shalf N f(0)\cr
&\ +\ N \Bigl[ { \textstyle -\ln\pi+a^2(\shalf\-\ln2)+\ln{2+a^2\over3a}
+(1\-a^2)({5\over12}\-\ln2) }\cr &\qquad\quad{\textstyle
+2\ln\shalf(1\+b)+{1\over3}(2\+a^2){1-b\over1+b}
+{1\over6}(1\-a^2)\bigl({1-b\over1+b}\bigr)^2\Bigr] +{\cal O}(1)}\cr }
\eqn\SPfinal $$
with $b^2\equiv3{2-a^2\over2+a^2}$ and a singular outcome for ${\cal
O}(1)$.
The one-loop correction yields
$$
\shalf\tr{}'\ln S_N''\ =\ {\textstyle \shalf\eta\ln N -2a\Bigl[1+\ln{\pi\over2}
-\ln{2-a^2\over a}-2\bigl(c^{-1}\arctan c -1\bigr) \Bigr] }
\eqn\oneloopfinal $$
where $c^2\equiv2{1-a^2\over2+a^2}$, which does not carry much meaning
in view of the singular classical contribution at the same order.

The comparison of these semiclassical ``results" with the exact
formula~\OPresult\ is poor.  Although the leading terms agree (as is
well-known) we have failed to reproduce the subleading contributions.
The logarithmic $N$-dependence may still be restored by taking
$\ln(x\-y)^2=\ln N^{-2}$ when $|x\-y|<N^{-1}$,
and by setting $\eta=-{5\over6}$.  The ${\cal O}(N)$ term is
(incorrectly) $a$-dependent and even at $g\=0\ (a^2\=1)$ does not
quite match up.  We do, however, expect corrections of {\cal O}(N) and
below from a proper regularization of the one-loop determinant
(see~\estimate), hence only the first line of eq.~\SPfinal\ can be
trusted anyway.  Presently, the first non-trivial subleading term of
$F_N(g)\-F_N(0)$ cannot be probed due to the divergencies, although it
is interesting to note that the logarithmic singularity at $a^2\=2$
already shows up correctly.
{\bf\chapter{Conclusions}}
We have formulated the collective field theory for hermitean matrix
models away from the large-$N$ limit.  By treating the Jacobian from
the change of variables from eigenvalues~$x_i$ to their
density~$\r(x)$ exactly, we arrived at a particular non-local and
non-polynomial theory of two real scalar fields, $\r$ and~$\la$.
Nevertheless, we were able to set up a loop expansion for the free
energy, at any $N\!<\!\infty$.  The classical eigenvalue density,
$\rh$, is determined by a simple integral equation whose large-$N$
limit becomes the usual airfoil equation.  Its action naturally
contains the entropy term for the Dyson gas of eigenvalues.  The
$\hbar$ expansion about the classical background yields sensible
Feynman rules, including somewhat formal expressions for the one-loop
determinant and the full propagator.  It has to be stressed that the
semiclassical (loop)~expansion is {\it not\/} identical to the
topological ($\N$)~expansion, because the ``equation of motion"
for~$\rh$ is not homogeneous in~$N$, thanks to the Jacobian.  We
studied the collective field theory in some detail up to the
three-loop level.

The attempt to develop our non-local field theory around $N\=\infty$
met with three difficulties.  First, the classical density~$\rh(x)$
can be expanded in~$\N$ only near the minimum of the matrix
potential~$V(x)$; elsewhere it behaves asymptotically as $e^{-NV(x)}$,
a strong non-perturbative (in~$\N$) signal.  The mixing of both
regions due to the defining integral equation destroys the expansion
everywhere; only the leading term $\rh\to\r_0$ is non-singular.
Second, UV singularities $\propto(\tr{\bf1})^k$ appear at $k$~loops
since the propagator is a delta-function to leading order.  Amazingly,
most of the divergent graphs cancel each other, except for the
connected diagrams with the maximal number of loops and no
self-contractions.  As a result, the free energy is not just expanded
in~$\N$ but also in~${\d(0)\over N\rh}$.  Third, the ensuing integrals
$\int\rh^{1-k}$ do not converge due to the exponentially small tail
of~$\rh$, producing IR divergencies.  A simple recipe to deal with
these problems is to restrict the eigenvalue space to a finite
interval and to implement a high-frequency cutoff.  Preliminary
answers were obtained here at the one-loop level.

Finally, a simple example was chosen to substantiate our abstract
discussion and to compare with known results obtained by the method of
orthogonal polynomials.  As anticipated, only the planar contribution
was reproduced entirely, but some features of the torus correction
appeared as well.

Although we focused on the zero-dimensional hermitean one-matrix
ensemble, the discussions in this paper should carry over mutatis
mutandis to multi-matrix models, the one-dimensional theory, and to
unitary matrix models.

It must be said that we did not yet achieve our goal of deriving a
practical algorithm which systematically computes $\N$ corrections in
the semiclassical scheme.  However, the necessary steps have been
taken, and the details of our failure are no more puzzling as they are
intriguing.  We certainly hope to gain further insight into this
outstanding problem in the future.

\ack
I wish to thank D.~Karabali, J.~Schiff and S.~Wadia for useful
discussions.
{\bf\Appendix{A}}
We give the basic notions of the infinite and finite Hilbert
transforms, including some useful properties.  Hilbert was the first
to notice the (skew-)reciprocity of
$$
u(x)\ =\ {1\over\pi}-\!\!\!\!\!\!\!\int_{-\infty}^\infty\!{dy\over x-y}\,v(y)
\quad,\qquad
v(x)\ =\ -{1\over\pi}-\!\!\!\!\!\!\!\int_{-\infty}^\infty\!{dy\over x-y}\,u(y)
\quad;
\eqn\reciprocity $$
$u$ is said to be conjugate to $v$.  The (infinite) Hilbert transform
$$
u(x)\ =\ \h_x[v]\ :=
\ {1\over\pi}-\!\!\!\!\!\!\!\int_{-\infty}^\infty\!{dy\over x-y}\,v(y)
\eqn\Hilbert $$
is related to Fourier integrals and to the Laplace transform and
received an extensive treatment in Titchmarsh's book~[\titchmarsh].
For $v\in L^{1+\e}$, $\h[v]$ exists almost everywhere and also belongs
to $L^{1+\e}$, so its inverse is
$$
v(x)\ \equiv\ \h^{-1}_x[u]\ =\ -\h_x[u] \quad.
\eqn\invHilbert $$
If $v\in L$, $\h[v]$ still exists almost everywhere but does not
necessarily belong to~$L$.  We may, however, set $\h[1]=0$.
Furthermore, it is easily seen that
$$
\h_x\bigl[v(ay\+b)\bigr]\ =\ {\textstyle{a\over|a|}}\,\h_{ax+b}\bigl[v(y)\bigr]
\quad.
\eqn\linearity $$

Some interesting properties of \h\ generalize to the {\it finite\/}
Hilbert transform,
$$
u(x)\ =\ \T_x[v]\ :=
\ {1\over\pi}-\!\!\!\!\!\!\int_a^b\!{dy\over x-y}\,v(y) \quad,
\eqn\finiteH $$
which plays an important role in aerodynamics and has been
investigated by a number of authors~[\sohngen, \tricomi].  Most useful
of those is Tricomi's convolution theorem,
$$
\T\bigl\{v_1\,\T[v_2] + v_2\,\T[v_1]\bigr\}\ =\ \T[v_1]\,\T[v_2]-v_1\,v_2
\quad,
\eqn\convolution $$
valid for $v_i\in L^{p_i}$ with $p_i>1$ and ${1\over p_1}+{1\over
p_2}<1$.  It is equivalent to the interchange formula
$$
-\!\!\!\!\!\!\int_a^b\!dy
-\!\!\!\!\!\!\int_a^b\!dz\,{v_1(y)\;v_2(z)\over(x-y)(y-z)}\ =
\ -\!\!\!\!\!\!\int_a^b\!dz
-\!\!\!\!\!\!\int_a^b\!dy\,{v_1(y)\;v_2(z)\over(x-y)(y-z)}\ -\ \pi^2
v_1(x)\,v_2(x)
\eqn\interchange $$
of Hardy and Poincar\'e (more generally, $v_1(y)\,v_2(z)\to
F(x,y,z)$).  Under the same conditions we also have Parseval's formula
$$
\int_a^b\!dx -\!\!\!\!\!\!\int_a^b\!dy\,{v_1(x)\,v_2(y)\over x-y}\
=\ \int_a^b\!dy -\!\!\!\!\!\!\int_a^b\!dx\,{v_1(x)\,v_2(y)\over x-y}
\eqn\Parseval $$
which may also be written as
$$
\int_a^b\!dx\;\Bigl( v_1\,\T[v_2] + v_2\,\T[v_1] \Bigr)\ =\ 0 \quad,
\eqn\skewadjoint $$
demonstrating the skew-adjoinedness of the \T\ (and \h) operations
with respect to the standard scalar product.  Another convenient
relation is
$$
\T'_x[v]\ =\ \T_x[v']
\eqn\Tprime $$
which allows us to commute a derivative past the transform, as
in~\finv.  All these relations, \convolution-\Tprime, hold for \h\ as
well.

Conversely, the linearity property~\linearity\ cannot hold for~\T;
rather
$$
\T_x[1]\ =\ -{1\over\pi} \ln\Bigl|{x-b\over x-a}\Bigr| \quad,
\eqn\Tofone $$
so that \convolution\ with $v_1\=1$ yields
$\T\bigl\{\T[v]\bigr\}\ne-v$, in contradistinction to~\h.  This
slightly complicates the inversion of~$\T$.  However, it is easily
calculated that
$$
{\textstyle
\T_x\bigl[{1\over\sqrt{(b-y)(y-a)}}\bigr]\ =\ 0 \quad,\qquad
\T_x\bigl[{\scriptstyle\sqrt{(b-y)(y-a)}}]\ =\ x-{a+b\over2} }
\eqn\Tsimple $$
on the interval $x\in(a,b)$.\foot{ A note of caution: outside $(a,b)$
the \rhs\ acquire an additional term of
$\pm\e\bigl[(x\-b)(x\-a)\bigr]^{\mp\half}$, respectively, with
$\e=\Theta(x\-b)-\Theta(a\-x)=\pm1$.} With these data a clever
application of the convolution theorem~\convolution\ then
establishes~[\tricomi]
$$
\T_x^{-1}[u]\ =\ {1\over\sqrt{(b-x)(x-a)}}\biggl[\gamma\ -\ {1\over\pi}
-\!\!\!\!\!\!\int_a^b \!{dy\over x-y}\;u(y)\;\sqrt{(b-y)(y-a)}\biggr]
\eqn\invT $$
where, using \Parseval\ and~\Tsimple,
$$
\gamma\ =\ {1\over\pi}\int_a^b \! dx\;\T_x^{-1}[u]
\eqn\norm $$
is an arbitrary normalization constant.  The conditions of the
convolution theorem restrict the validity of the inversion~\invT\
to~$u\in L^{{4\over3}+\e}$, but the uniqueness of the kernel~\Tsimple\
holds in~$L^{1+\e}$.  For $u\in L^{1+\e}$ the \rhs\ may not be in the
same class; on the other side, any inverse of~\T\ in~$L^{1+\e}$ must
be of the form~\invT.  Furthermore, given $u\in L^{2+\e}$, $\T[u]$ is
necessarily orthogonal to the zero mode,
$$
\int_a^b\!dx\,{\textstyle{1\over\sqrt{(b-x)(x-a)}}}\;\T_x[u]\ =\ 0
\eqn\orthogonal $$
by virtue of the skew-adjoinedness~\skewadjoint.  This allows for
alternative representations, \eg\
$$
\T_x^{-1}[u]\ =\ {\gamma'\over\sqrt{(b-x)(x-a)}}\ -\ {1\over\pi}
-\!\!\!\!\!\!\int_a^b\!
{dy\over x-y}\;u(y)\;{\sqrt{(b-x)(x-a)}\over\sqrt{(b-y)(y-a)}}
\eqn\alternative $$
which also goes to~\invHilbert\ when $(a,b)\to(-\infty,\infty)$.

Unfortunately, our nonlinear extension of the Hilbert transform,
$$
u(x)\ =\ \h_x[v] - \beta{v'\over v}(x)
\eqn\nonlinear $$
does not seem to have been investigated in the literature.
{\bf\Appendix{B}}
In the case of a gaussian ensemble ($V\=\half x^2$) of hermitean
random matrices the free energy~$F_N$ for any~$N$ can be found {\it
exactly\/}, with orthogonal polynomial techniques.  As usual~[\bessis]
$$
e^{-F_N}\ :=\ \biggl[\prod_{i=1}^N \int\! dx_i\biggr] \; e^{-\half
N \sum_i x_i^2}\;\prod_{i<j}(x_i-x_j)^2 \ =\ N!\,\prod_{k=0}^{N-1} h_k
\eqn\gaussZ $$
where $h_k$ is the norm of the $k$th monic orthogonal polynomial,
$P_k(x)=x^k+\ldots$.  For our quadratic potential the $P_k$ are just
proportional to the Hermite polynomials,
$$
P_k(x)\ =\ {\scriptstyle\sqrt{2N}^{-k}}
\,H_k\bigl(x{\scriptstyle\sqrt{N/2}}\bigr)\quad,
\eqn\hermite $$
whose norms are well-known,
$$
h_k\ =\ \medint_{-\infty}^{+\infty}
dx\;e^{-\half N x^2}\,P_k(x)^2\ =\ N^{-k-\half}\,k!\,\sqrt{2\pi}\quad.
\eqn\norms $$
Hence, we find
$$
e^{-F_N}\ =\ N^{-N^2/2}\;(2\pi)^{N/2}\,\prod_{k=1}^N k!
\eqn\exactZ $$
which is the exact result.

We go on to expand
$$
F_N\ =\ \shalf N^2 \ln N\ -\ \shalf N \ln 2\pi
\ -\ {\textstyle\sum_1^N} (N\-k\+1)\ln k
\eqn\exactF $$
in powers of~$\N$, with the help of Stirling's asymptotic
formula\foot{ $B_{2k}$ denotes the Bernoulli numbers:
$B_2\={1\over6}$, $B_4\=-{1\over30}$, etc.}
$$
\eqalign{
\ln\Gamma(z)\ &=\ (z\-\shalf)\ln z - z + \shalf\ln2\pi +
{\textstyle \sum_1^{s-1} {B_{2k}\over2k(2k-1)}\,z^{1-2k} } + R_s(z)
\cr |R_s(z)|\ &<\ {\textstyle{|B_{2s}|\over 2s(2s-1)\,|z|^{2s-1}\,
\cos^{2s-1}(\half\arg z)}} \cr}
\eqn\stirling $$
for large values of $|z|$ as well as
$$
\sum_{k=1}^N k\,\ln k\ =\ \int_0^{N+1}\!\!dt\,\ln\Gamma(t)
\ +\ \shalf(N\+1)(N\-\ln2\pi)
\eqn\sumklnk $$
$$
\int_p^{p+1}\!\!dt\,\ln\Gamma(t)\ =\ p\,\ln p\ -\ p\ +\ \shalf\ln2\pi \quad.
\eqn\intlngamma $$
The $N^2\ln N$ terms cancel, and after collecting all the pieces we
arrive at the asymptotic series
$$
F_N\ \simeq\ {3\over4}N^2 - N\ln N
+ (1\-\ln2\pi)N - {5\over12}\ln N + c_0 - \sum_{k=2}^\infty
{B_{2k}\;N^{2-2k}\over 2k(2k\-1)(2k\-2)}
\eqn\exactexp $$
where
$$
\eqalign{
c_0\ &\simeq\ - {\textstyle \shalf\ln2\pi-{1\over12}+\sum_2^{p-1}k\ln k }\cr
&\qquad+ {\textstyle {1\over4}p^2-\shalf(p^2\-p\+{1\over6})\ln p
+\sum_2^\infty {B_{2k}\;\,p^{2-2k}\over 2k(2k-1)(2k-2)} }\cr &\ =
\ -0.7535173895042\ldots \cr}
\eqn\universal $$
is a universal constant, defined ever more accurately for ever larger
values of the cut-off parameter~$p$.  As expected on general grounds,
this is an expansion in~${\textstyle{1\over N^2}}$, except for a few
terms at low orders.
{\bf\Appendix{C}}
We outline a rough estimate of the determinant of
the fluctuation operator~\fluctop, regulated by an IR cutoff~$L$ and a
UV mode cutoff~$M$.  In effect, we restrict the eigenvalue range to
$[-{L\over2},+{L\over2}]$ with open boundary conditions and convert to
Fourier modes
$$
\r(x)\ =\ {\textstyle{1\over L}}\sum_{k\in{\bf Z}} e^{2\pi ikx/L}\;\r_k
\eqn\rhomodes $$
before setting $\r_k\=0$ for $|k|>M$.

The Fourier transform of $S''_N(x,y)$ reads
$$
S''_{kl}\ =\ \pmatrix{-N^2 f_{kl} & -i\d_{k+l,0} \cr -i\d_{k+l,0} &
\N\bigl[\rh_{k+l}-\rh_k\rh_l\bigr] \cr}
\eqn\fluctopfourier $$
which, for the simple case of $V\=\shalf x^2$ and after reinstating
$f(z)\to\ln z^2$, can be evaluated explicitly,
$$
\eqalign{
\rh_k\ &=\ {\textstyle{L\over2\pi k}\,{\rm J}_1\bigl({4\pi k\over L}\bigr)}\cr
f_{kl}\ &=\ {\textstyle{-2\over\pi k}\bigl({\rm si}(\pi k)+{\pi\over2}\bigr)
\,\d_{k+l,0}}\ \equiv\ f_k\,\d_{k+l,0}\cr}
\eqn\fluctopmodes $$
with ${\rm si}(x)\equiv-\int_x^\infty{\sin t\over t}$.  By some
algebraic identities we rewrite
$$
\det{}'S''\ =\ \det{}'\bigl[\d_{k+l,0} -N f_k(\rh_{k+l}-\rh_k\rh_l)\bigr]\ =
\ \det\bigl[ (\d_{k+l,0}-\d_{k,0}\d_{l,0})-N\sqrt{f_k}\rh_{k+l}\sqrt{f_l}\bigr]
\eqn\newdet $$
where the prime means omission of the zero mode ($k\=0$), and we have
used $\rh_0\=1$.

For large values of $|k|$, one has
$$
\rh_k\ \approx
\ {\textstyle\sqrt{{L^3\over8\pi^4}}|k|^{-3/2}\cos({4\pi|k|\over L}
-{3\pi\over4})} \qquad,\qquad f_k\ \approx\ -{1\over|k|}
\eqn\largemodes $$
so that the matrix elements for~\newdet\ fall off as $|k\+l|^{-5/2}$
perpendicular to the diagonal.  We can make a rough estimate of
$\det{}'S''$ by boldly taking into account only the diagonal elements,
\ie\
$$
\det{}'S''\Big|_{\rm reg}\ \approx
\ -Nf_0 \prod_{k=1}^M \bigl(1+{\textstyle{N\over k}}\bigr)^2
\ =\ 2N {\textstyle{M+N\choose N}^2}
\eqn\estimate $$
since $f_0\=-2$.  Let us now choose a UV cutoff proportional to~$N$,
say $M\=\a N\+\beta$.  Via Stirling's formula then
$$
\shalf\tr{}'\ln S''\Big|_{\rm reg}
\ \approx\ \bigl[(\a\+1)\ln(\a\+1)-\a\ln\a\bigr]\,N\ +\ {\cal O}(\ln N)\quad.
\eqn\rough $$
More accurate estimates have to consider some sub-diagonals as well.
\vfil \eject
\refout
\bye